\documentclass[A4,useAMS,usenatbib,onecolumn]{mn2e}

\usepackage{amsmath}
\usepackage{amssymb}

\newcommand{\be}{\begin{equation}}
\newcommand{\ee}{\end{equation}}
\newcommand{\ba}{\begin{eqnarray}}
\newcommand{\ea}{\end{eqnarray}}

\newcommand{\rmd}{\textrm{d}}

\usepackage{aas_macros}
\usepackage{graphicx}
\usepackage{bm}
\usepackage{natbib}
\usepackage{mathrsfs}
\usepackage{color}
\usepackage[mediumspace,Gray,squaren]{SIunits}

\title[tSZ-CIB correlation]{Modelling the correlation between the thermal Sunyaev Zel'dovich effect and the cosmic infrared background}
\author[G. E. Addison et al.]{G. E. Addison$^{1}$\thanks{E-mail: graeme.addison@astro.ox.ac.uk}, J. Dunkley$^{1}$ and D. N. Spergel$^{2}$\\
$^{1}$Sub-department of Astrophysics, University of Oxford, Denys Wilkinson Building, Keble Road, Oxford OX1 3RH, UK\\
$^{2}$Department of Astrophysical Sciences, Peyton Hall, Princeton University, Princeton, NJ USA 08544}

\begin{document}

\date{Accepted xxx. Received xxx; in original form xxx}

\pagerange{\pageref{firstpage}--\pageref{lastpage}} \pubyear{2012}

\maketitle

\label{firstpage}

\begin{abstract}
We show how the correlation between the thermal Sunyaev Zel'dovich effect (tSZ) from galaxy clusters and dust emission from cosmic infrared background (CIB) sources can be calculated in a halo model framework. Using recent tSZ and CIB models, we find that the size of the tSZ$\times$CIB cross-correlation is approximately 20 per cent at 150~GHz. The contribution to the total angular power spectrum is around $-2$~$\mu$K$^2$ at $\ell=3000$, however, this value is uncertain by a factor of two to three, primarily because of CIB source modelling uncertainties. We expect the large uncertainty in this component to degrade upper limits on the kinematic Sunyaev Zel'dovich effect (kSZ), due to similarity in the frequency dependence of the tSZ$\times$CIB and kSZ across the frequency range probed by current Cosmic Microwave Background missions. We also find that the degree of tSZ$\times$CIB correlation is higher for mm$\times$sub-mm spectra than mm$\times$mm, because more of the sub-mm CIB originates at lower redshifts ($z\lesssim2$), where most tSZ clusters are found.
\end{abstract}


\section{Introduction}
The ability to connect the luminous baryonic matter our telescopes see with the underlying dark matter density field is fundamental to our understanding of large-scale structure and galaxy evolution. Combining the `halo model' with a Halo Occupation Distribution (HOD) prescription achieves this through associating luminous sources with collapsed dark matter haloes, the peaks of the density distribution (e.g., \citealt{peacock/smith:2000}; \citealt{scoccimarro/etal:2001}; \citealt{berlind/weinberg:2002}; \citealt{cooray/sheth:2002}; also \citealt{bond:1996} and references therein). This framework has been successfully applied to interpret the clustering properties of several source populations, including local galaxies from the Sloan Digital Sky Survey (SDSS), luminous red galaxies (LRGs), and high-redshift Lyman-break galaxies, using measurements of the angular correlation function \citep[e.g.,][]{zehavi/etal:2004,ouchi/etal:2005,blake/etal:2008,zheng/etal:2009,zehavi/etal:2011}.

A natural application of the halo model is to understand the correlation between two different source populations, each tracing the underlying dark matter but having different dependence on host halo properties \citep[e.g.,][]{cooray/sheth:2002}. In this work we present such a treatment for the angular power spectrum arising from the correlation between unresolved clusters contributing to the thermal Sunyaev Zel'dovich effect \citep[tSZ;][]{sunyaev/zeldovich:1970}, and the dusty sources that make up the cosmic infrared background \citep[CIB; e.g.,][]{puget/etal:1996}. Along with radio sources, these two populations are currently of interest as foregrounds in Cosmic Microwave Background (CMB) temperature anisotropy analysis \citep[e.g.,][]{white/majumdar:2003,righi/etal:2008}. The tSZ effect arises from the Compton scattering of CMB photons by hot electrons in massive galaxy clusters, while the CIB sources are understood to be mostly high-redshift ($z\sim1-4$) galaxies undergoing active, dust-enshrouded star formation \citep[e.g.,][]{bond/etal:1986,bond/etal:1991c,hughes/etal:1998,blain/etal:1999,draine:2003}. Absorption of the starlight by the dust grains and subsequent thermal re-emission leads to strong rest-frame far-infrared emission, and cosmic expansion stretches these photons' wavelengths so they are detected in the mm-bands of contemporary CMB missions including the South Pole Telescope \citep[SPT;][]{hall/etal:2010}, the Atacama Cosmology Telescope \citep[ACT;][]{fowler/etal:2010}, and the \emph{Planck} satellite \citep{planckcib:2011}.

The size of the tSZ$\times$CIB cross-correlation provides physical information about the extent to which the CIB sources are associated with the clusters that contribute to the tSZ signal. Understanding how and why star formation in massive systems has evolved over cosmic time \citep[e.g,][]{chapman/etal:2005,pannella/etal:2009,magnelli/etal:2011} is key to understanding galaxy evolution. In addition, \citeauthor{reichardt/etal:2012} (2012 -- hereafter R12) have recently found that uncertainty in the tSZ$\times$CIB power significantly degrades constraints on the kinematic Sunyaev Zel'dovich effect (kSZ) from SPT data. The kSZ arising from peculiar motion of galaxy clusters has recently been detected \citep{hand/etal:2012}, however improving constraints on the kSZ power from patchy reionization -- in which CMB photons scatter off electrons in expanding ionized gas bubbles \citep{knox/etal:1998} -- is of great interest for constraining the duration of reionization and thus star formation in the early universe. For these reasons, a model for the tSZ$\times$CIB power is required, and we choose a halo model framework since, as stated, it provides a natural basis for relating two different tracers of the matter field, and, furthermore, it has already been used to study the tSZ and CIB anisotropies.

We note that a similar halo model approach has recently been used to model the cross-correlation of the tSZ effect with the integrated Sachs--Wolfe effect \citep{taburet/etal:2011}, and the distribution of galaxy clusters \citep{fang/etal:2012}.

In Section 2 we collate the halo model equations for the tSZ and CIB power spectra and present expressions describing the tSZ$\times$CIB cross-correlation. In Sections 3 and 4 we combine recent models for the tSZ and CIB and calculate the tSZ$\times$CIB contribution as a function of frequency and angular scale. The uncertainties in current models and the role of upcoming data sets are discussed in Section 5 and conclusions follow in Section 6.

For the results presented in Section 4 we adopt a flat, $\Lambda$CDM cosmology, with $h=H_0/100~$km~s$^{-1}$~Mpc$^{-1}=0.70$, $\Omega_{\rm m}=0.27$, $\Omega_{\rm b}=0.045$, and $\sigma_8=0.80$, consistent with WMAP-7 results \citep{komatsu/etal:2011}.

\section{Model}
Analysis of mm CMB data is typically performed using thermodynamic temperature units, while in the sub-mm, where the CIB is dominant, it is more natural to work with units of flux density. The equations in this section contain a mix of units, however the plots in Section 4 are made in either flux or temperature units. The conversion from thermodynamic temperature to flux density requires multiplying by $\partial B_{\nu}(T)/\partial T|_{T=T_{\rm CMB}}$, where
\be
\frac{\partial B_{\nu}}{\partial T}(T)=\frac{2k_{\rm B}\nu^2}{c^2}\frac{x^2e^x}{\left(e^x-1\right)^2},
\ee
with $x=h_{\rm p}\nu/k_{\rm B}T$ and the present-day CMB temperature $T_{\rm CMB}=2.725~\rm K$. Note that converting between units requires an integral across the bandpass filter for the finite bands of real experiments.

\subsection{CIB power}
The angular power spectrum of unresolved CIB sources at multipole moment $\ell$ from correlating two maps at frequency $\nu_1$ and $\nu_2$ may be written as
\be
C_{\ell,\nu_1\nu_2}^{\rm CIB}=C_{\ell,\nu_1\nu_2}^{\rm CIB-clust}+C_{\nu_1\nu_2}^{\rm CIB-P}.
\ee
where the clustered component, $C_{\ell}^{\rm CIB-clust}$, arises from the correlation of images of different sources in the two maps, while the Poissonian shot noise component, $C^{\rm CIB-P}$, is due to the finite number of sources in any given observed field, and may be thought of as arising from the correlation of images of the same source in the two maps. For the range of angular scales considered ($\sim$arcminute and larger), the CIB sources are effectively point sources, and so the CIB shot noise is scale-independent.

In the small-sky limit \citep{limber:1953}, the clustered power may be written as \citep[e.g.,][]{knox/etal:2001,tegmark/etal:2002}:
\be
\label{V09eqn}
C_{\ell,\nu_1\nu_2}^{\rm CIB-clust}=\int \textrm{d}z\left(\frac{\textrm{d}V_{\rm c}}{\textrm{d}z}(z)\right)^{-1}\frac{\textrm{d}I^{\rm CIB}_{\nu_1}}{\textrm{d}z}(z)\frac{\textrm{d}I^{\rm CIB}_{\nu_2}}{\textrm{d}z}(z)P_{\rm gal}\left(k=\ell/\chi(z),z\right),
\ee
where $\chi$ is the comoving distance, $\textrm{d}V_{\rm c}/\textrm{d}z=\chi^2\textrm{d}\chi/\textrm{d}z$ the comoving volume element, $\rmd I^{\rm CIB}/\rmd z$ the redshift distribution of the CIB intensity (sometimes called $\rmd S/\rmd z$), and $P_{\rm gal}$ is the three-dimensional power spectrum of the dusty galaxies (and any other sources) that make up the CIB. 

We write $P_{\rm gal}$ as the sum of one- and two-halo terms, corresponding to the contribution from pairs of sources lying in the same dark matter halo, and different haloes, respectively \citep{cooray/sheth:2002}:
\be
P_{\rm gal}(k,z)=P_{\rm gal}^{\rm 1h}(k,z)+P_{\rm gal}^{\rm 2h}(k,z),
\ee
where
\be
P_{\rm gal}^{\rm 1h}(k,z)=\int\rmd M\,\frac{\rmd n_{\rm h}}{\rmd M}(M,z)\frac{\langle N_{\rm gal}\left(N_{\rm gal}-1\right)\rangle}{n_{\rm gal}^2(z)}u_{\rm gal}^{2}(k,M,z)
\ee
and
\be
P_{\rm gal}^{\rm 2h}(k,z)=P_{\rm{DM}}(k,z)\bigg(\int\rmd M\,\frac{\rmd n_{\rm h}}{\rmd M}(M,z)\,b_{\rm h}(k,M,z)
\frac{\langle N_{\rm gal}\rangle}{n_{\rm gal}(z)} \,u_{\rm gal}(k,M,z)\bigg)^2.
\ee
The integral is over halo mass; $\rmd n_{\rm h}/\rmd M$ is the halo mass function, $N_{\rm gal}(M,z)$ the number of CIB sources hosted by a halo, $u_{\rm gal}$ the Fourier transform of the spatial distribution of the sources within their halo, and $b_{\rm h}$ the halo bias relative to the dark matter power spectrum, $P_{\rm DM}$, such that the halo and dark matter power spectra are related by
\be
P_{\rm h}(k,M,z)=b_{\rm h}^2(k,M,z)\,P_{\rm DM}(k,z).
\ee
The comoving CIB source number density, $n_{\rm gal}$, is given by
\be
n_{\rm gal}(z)=\int\rmd M\,\frac{\rmd n_{\rm h}}{\rmd M}(M,z)\langle N_{\rm gal}\rangle,
\ee
and the angled brackets denote averaging over all haloes of mass $M$ at redshift $z$.

Note that equation (\ref{V09eqn}) assumes that the spectral properties of the CIB sources are not dependent on properties of their host haloes. To make our approach more general, and to facilitate writing down the tSZ$\times$CIB cross-correlation term later, we allow the CIB flux from a source at a given redshift to depend on halo mass \citep[as in, e.g.,][]{righi/etal:2008,sehgal/etal:2010,shang/etal:2012}. This requires weighting each integral over halo mass with the mean source flux, $\langle S_{\nu}(M,z)\rangle$, yielding (omitting dependence on angular scale, redshift and halo mass for brevity)
\be
\label{P_1h}
P_{\rm gal,\nu_1\nu_2}^{\prime\rm 1h}=\int\rmd M\,\frac{\rmd n_{\rm h}}{\rmd M}\frac{\langle S_{\nu_1}\rangle\langle S_{\nu_2}\rangle\langle N_{\rm gal}\left(N_{\rm gal}-1\right)\rangle}{s_{\nu_1}s_{\nu_2}}\,u_{\rm gal}^{2}
\ee
and
\be
\label{P_2h}
P_{\rm gal,\nu_1\nu_2}^{\prime\rm 2h}=P_{\rm{DM}}\int\rmd M_1\,\frac{\rmd n_{\rm h}}{\rmd M_1}\,b_{\rm h}\frac{\langle S_{\nu_1}\rangle\langle N_{\rm gal}\rangle}{s_{\nu_1}} \,u_{\rm gal}\int\rmd M_2\,\frac{\rmd n_{\rm h}}{\rmd M_2}\,b_{\rm h}\frac{\langle S_{\nu_2}\rangle\langle N_{\rm gal}\rangle}{s_{\nu_2}} \,u_{\rm gal},
\ee
where
\be
s_{\nu}(z)=\int\rmd M\,\frac{\rmd n_{\rm h}}{\rmd M}(M,z)\langle S_{\nu}(M,z)\rangle\langle N_{\rm gal}\rangle.
\ee
Substituting into equation (\ref{V09eqn}), and recognizing that $\rmd I_{\nu}^{\rm CIB}/\rmd z=s_{\nu}\rmd V_{\rm c}/\rmd z$, gives
\be
C_{\ell,\nu_1\nu_2}^{\rm CIB}=C_{\ell,\nu_1\nu_2}^{\rm CIB-1h}+C_{\ell,\nu_1\nu_2}^{\rm CIB-2h}+C_{\nu_1\nu_2}^{\rm CIB-P},
\ee
with the following expressions for the one- and two-halo terms:
\be
C_{\ell,\nu_1\nu_2}^{\rm CIB-1h}=\int\rmd z \frac{\rmd V_{\rm c}}{\rmd z}\int \rmd M\,\frac{\rmd n_{\rm h}}{\rmd M}\,\langle S_{\nu_1}\rangle\,\langle S_{\nu_2}\rangle\langle N_{\rm gal}\left(N_{\rm gal}-1\right)\rangle\,u_{\rm gal}^{2},
\ee
and
\be
C_{\ell,\nu_1\nu_2}^{\rm CIB-2h}=\int\rmd z \frac{\rmd V_{\rm c}}{\rmd z}P_{\rm DM}\int \rmd M_1\,\frac{\rmd n_{\rm h}}{\rmd M_1}\,b_{\rm h}\,\langle S_{\nu_1}\rangle\,\langle N_{\rm gal}\rangle\,u_{\rm gal}\int \rmd M_2\,\frac{\rmd n_{\rm h}}{\rmd M_2}\,b_{\rm h}\,\langle S_{\nu_2}\rangle\,\langle N_{\rm gal}\rangle\,u_{\rm gal}.
\ee
The CIB shot noise contribution is given by \citep[e.g.,][]{scott/white:1999}
\be
C_{\nu}^{\rm CIB-P}=\int\rmd S_{\nu}\frac{\rmd N}{\rmd S_{\nu}}S_{\nu}^2,
\ee
so in terms of the halo model quantities defined above we have
\be
C_{\nu_1\nu_2}^{\rm CIB-P}=\int\rmd z\frac{\rmd V_{\rm c}}{\rmd z}\int \rmd M\,\frac{\rmd n_{\rm h}}{\rmd M}\,\langle S_{\nu_1} S_{\nu_2}\rangle\langle N_{\rm gal}\rangle.
\ee
Note that the shot noise depends on $\langle S_{\nu_1} S_{\nu_2}\rangle$ whereas the one-halo power depends on $\langle S_{\nu_1}\rangle\langle S_{\nu_2}\rangle$. If the framework described here is to be used to perform joint fits to the shot noise and clustered power, the distribution of CIB source flux at fixed $M$ and $z$ must therefore be considered.

Further modifications to these equations, such as distinguishing between the first, `central', source occupying a halo and subsequent, `satellite', sources \citep[e.g.,][]{berlind/etal:2003,kravtsov/etal:2004,zheng/etal:2005,tinker/etal:2010b}, may easily be made.

The CIB intensity (one-point function), $I_{\nu}^{\rm CIB}$, is simply the integral of $\rmd I_{\nu}^{\rm CIB}/\rmd z$, and is therefore given by
\be
\label{I_CIB}
I_{\nu}^{\rm CIB}=\int\rmd z\frac{\rmd V_{\rm c}}{\rmd z}\int\rmd M\,\frac{\rmd n_{\rm h}}{\rmd M}\langle S_{\nu}\rangle\langle N_{\rm gal}\rangle.
\ee
Note that the clustered power, shot noise and CIB intensity may all have quite different redshift dependence, due to the different factors appearing in the integrals in these equations. This point is demonstrated in Section 4.

\subsection{tSZ power}

In the halo model, the tSZ angular power spectrum from correlating two maps at frequency $\nu_1$ and $\nu_2$ is
\be
C_{\ell,\nu_1\nu_2}^{\rm tSZ}=C_{\ell,\nu_1\nu_2}^{\rm tSZ-1h}+C_{\ell,\nu_1\nu_2}^{\rm tSZ-2h}.
\ee
We follow previous studies \citep[e.g.,][]{komatsu/seljak:2002,shaw/etal:2009,battaglia/etal:2010} in treating each dark matter halo as hosting a single, extended source (electron population) for the purposes of contributing to the tSZ power. Unlike for the CIB sources, the one-halo term and the Poisson shot noise are then the same thing (which we will refer to as the one-halo term). The inter-cluster two-halo term has been neglected in existing mm-band CMB analysis, since, due to the scarcity of massive clusters, it is important only for $\ell\lesssim300$ \citep{komatsu/kitayama:1999}, and on these scales the primary CMB anisotropy is completely dominant.

Again invoking the small-sky approximation, we write the one- and two-halo tSZ components as \citep{komatsu/kitayama:1999,komatsu/seljak:2002}:
\be
C_{\ell,\nu_1\nu_2}^{\rm tSZ-1h}=g_{\nu_1}g_{\nu_2}\,T^2_{\rm CMB}\int\rmd z\frac{\rmd V_{\rm c}}{\rmd z}(z)\int\rmd M\,\frac{\rmd n_{\rm h}}{\rmd M}(M,z)\,\tilde{y}^2_{\ell}(M,z)
\ee
and
\be
C_{\ell,\nu_1\nu_2}^{\rm tSZ-2h}=g_{\nu_1}g_{\nu_2}\,T^2_{\rm CMB}\int\rmd z\frac{\rmd V_{\rm c}}{\rmd z}(z)\,P_{\rm DM}\large(k=\ell/\chi(z),z\large)\left(\int\rmd M\,\frac{\rmd n_{\rm h}}{\rmd M}(M,z)\,b_{\rm h}(k,M,z)\,\tilde{y}_{\ell}(M,z)\right)^2,
\ee
where $g_{\nu}$ is the spectral function of the tSZ effect, given, neglecting relativistic corrections, by
\be
g_{\nu}=\left(x\frac{e^x+1}{e^x-1}-4\right)\textrm{, with }x=\frac{h_{\rm p}\nu}{k_{\rm B}T_{\rm CMB}},
\ee
and $\tilde{y}_{\ell}$ is the two-dimensional Fourier transform of the Compton Y-parameter, $y_{\rm 3D}$ \citep{komatsu/seljak:2002}:
\be
\tilde{y}_{\ell}=\frac{4\pi r_{\rm s}}{\ell_{\rm s}^2}\int\rmd x\,x^2\,y_{\rm 3D}(x)\frac{\textrm{sin}(\ell x/\ell_{\rm s})}{\ell x/\ell_{\rm s}};
\ee
here $x=r/r_{\rm s}$, where $r$ is radius from the centre of the (assumed spherically symmetric) halo, $r_{\rm s}$ is a scale radius, and $\ell_{\rm s}\equiv D_A/r_{\rm s}$, with $D_A=\chi/(1+z)$ the angular diameter distance to redshift $z$.

Note that $g_{\nu}<0$ for $\nu\lesssim217$ GHz. This means that, for certain choices of $\nu_1$ and $\nu_2$, the tSZ power is \emph{negative}. This can also be the case for the tSZ$\times$CIB contribution.

The integrated Compton Y-parameter from a single cluster (integrated over the projected area of the cluster on the sky) is given by
\be
Y_{\rm SZ}=\frac{4\pi r_{\rm s}^3}{D_A^2}\int \rmd x\,x^2 y_{\rm 3D}(x).
\ee
Summing the contribution to the total tSZ intensity from all clusters we then have
\be
\label{I_tSZ}
I_{\nu}^{\rm tSZ}=g_{\nu}T_{\rm CMB}\int\rmd z\frac{\rmd V_{\rm c}}{\rmd z}\int\rmd M\,\frac{\rmd n_{\rm h}}{\rmd M}\,Y_{\rm SZ}.
\ee

\subsection{tSZ$\times$CIB power}

We now apply the halo model to describe the tSZ$\times$CIB cross-correlation. There is a one-halo term arising from the correlation of CIB sources with the tSZ from their host halo, but there is also a two-halo term arising from the correlation between the CIB sources in one halo and the tSZ from another. Even if we consider the extreme limit in which there is no star formation and no CIB sources in those haloes massive enough to make a non-negligible contribution to the tSZ, this second term would still exist, provided there was some overlap between the redshift distribution of the CIB sources and tSZ haloes. This is a key prediction of this work. On large angular scales, where the two-halo term dominates, the tSZ$\times$CIB power is therefore relatively insensitive to the astrophysical processes that affect star formation in dense massive haloes.

We write the total tSZ$\times$CIB power as
\be
C_{\ell,\nu_1\nu_2}^{\rm tSZ\times CIB}=C_{\ell,\nu_1\nu_2}^{\rm tSZ\times CIB-1h}+C_{\ell,\nu_1\nu_2}^{\rm tSZ\times CIB-2h}.
\ee
In terms of the quantities defined in Sections 2.1 and 2.2, the one- and two-halo terms are
\be
C_{\ell,\nu_1\nu_2}^{\rm tSZ\times CIB-1h}=T_{\rm CMB}\int\rmd z\frac{\rmd V_{\rm c}}{\rmd z}\int \rmd M\,\frac{\rmd n_{\rm h}}{\rmd M}\,\tilde{y}_{\ell}\bigg(g_{\nu_1}\langle S_{\nu_2}\rangle+g_{\nu_2}\langle S_{\nu_1}\rangle\bigg)\langle N_{\rm gal}\rangle\,u_{\rm gal}
\ee
and
\be
C_{\ell,\nu_1\nu_2}^{\rm tSZ\times CIB-2h}=T_{\rm CMB}\int \rmd z\frac{\rmd V_{\rm c}}{\rmd z}P_{\rm DM}\int \rmd M_1\,\frac{\rmd n_{\rm h}}{\rmd M_1}\,b_{\rm h}\,\tilde{y}_{\ell}\int \rmd M_2\,\frac{\rmd n_{\rm h}}{\rmd M_2}\,b_{\rm h}\bigg(g_{\nu_1}\langle S_{\nu_2}\rangle+g_{\nu_2}\langle S_{\nu_1}\rangle\bigg)\langle N_{\rm gal}\rangle\,u_{\rm gal}.
\ee

The fact that, in the non-relativistic limit, the tSZ spectral dependence is not coupled to redshift or halo mass leads to these cross-terms being simpler than would in general be the case for two source populations. While equations 26 and 27 include the tSZ and CIB contributions from both $\nu_1$ and $\nu_2$, an interesting case to consider is that of two widely spaced frequencies, such that one of the two terms in parentheses in equations 26 and 27 is negligible due to the different frequency dependence of the tSZ and CIB. Constraints from cross-spectra at widely-spaced frequencies may actually provide the best opportunity for detecting and constraining the tSZ$\times$CIB term. This idea is discussed further in Sections 4.2 and 5.2.

Using this formalism, the tSZ$\times$CIB cross-correlation power requires no additional functions or parameters over those used to characterize the tSZ$\times$tSZ and CIB$\times$CIB power individually. This is an advantage of the halo model approach over the template-based approach used to describe the tSZ and CIB power in R12, which requires the inclusion of a correlation coefficient $\xi$ to specify the tSZ$\times$CIB cross-power. Furthermore, the frequency and angular scale dependence of $\xi$ may not be negligible, meaning fitting a single correlation coefficient may not be sufficient to describe the tSZ$\times$CIB power (see Section 4.5).

\section{Example calculation}

Having laid out an analytic halo model framework for the tSZ$\times$CIB power, we now use recent models for the CIB and tSZ to perform calculations at frequencies probed by current arcminute-resolution mm and sub-mm telescopes: ACT, SPT, \emph{Planck}'s High Frequency Instrument, and \emph{Herschel}'s Spectral and Photometric Imaging Receiver (SPIRE) instrument. We focus on 150~GHz as the band primarily used to constrain small-scale CMB anisotropy power by ACT and SPT \citep[e.g.,][]{hall/etal:2010,fowler/etal:2010,keisler/etal:2011,dunkley/etal:2011}, and cross-correlations with higher frequencies as an opportunity to potentially improve constraints on the tSZ$\times$CIB amplitude and angular scale dependence. Specifically, we perform calculations using bandpass filters from the following frequency channels: SPT's 150~GHz (2100~$\mu$m), \emph{Planck}'s 857, 545, 353 and 217 GHz (350, 550, 840 and 1380~$\mu$m), and SPIRE's 1200 GHz (250~$\mu$m). Predictions are also made for the ACT 148 and \emph{Planck} 143~GHz channels.

\subsection{CIB model}

For the CIB clustering power we use the model of \citeauthor{xia/etal:2012} (2012 -- hereafter X12). This work uses the CIB source model developed in \cite{granato/etal:2004} and \citeauthor{lapi/etal:2011} (2011 -- hereafter L11). Three populations of sources are considered: spiral galaxies, starbursts (these two populations contribute to the CIB for $z\lesssim2$), and proto-spheroids (the dominant CIB sources at high redshift, which do not exist at $z<1$). X12 perform a joint fit to angular power spectra from \emph{Herschel}/SPIRE \citep{amblard/etal:2011}, \emph{Planck} \citep{planckcib:2011} and SPT \citep{shirokoff/etal:2011}, using a simple HOD model based on \cite{zheng/etal:2005}. There are two free parameters in the model: a minimum halo mass for hosting a proto-spheroid source, $M_{\rm min}$, and the power-law index, $\alpha_{\rm sat}$, which describes the occupation of massive haloes with additional proto-spheroids. The results in Section 4 are shown for $\log(M_{\rm min}/M_{\sun})=12.24$ and $\alpha_{\rm sat}=1.81$, the means of the marginalized parameter distributions obtained by X12. We find that the statistical uncertainty from varying these two parameters is not significant compared to other uncertainties in the tSZ$\times$CIB power (see Section 5.1).

Note that the X12 model does not include any dependence of CIB source spectral properties (luminosity, SED, etc.) on host halo mass. Consequently, the angular power spectrum is evaluated simply using equation (3) and the $\rmd I_{\nu}^{\rm CIB}/\rmd z$ predictions from L11.

\subsection{tSZ model}

For the tSZ we use the model described in \citeauthor{efstathiou/migliaccio:2012} (2012 -- hereafter EM12), which is based on the halo model of \cite{komatsu/seljak:2002}, but with the empirical electron pressure profiles from recent X-ray measurements of the Representative XMM-Newton Cluster Structure Survey cluster sample \citep{arnaud/etal:2010}. The model has two free parameters: an overall amplitude, $A$ (such that $C_{\ell}^{\rm tSZ}\propto A$), and an evolution parameter, $\epsilon$, which describes departures from self-similar evolution of the electron pressure profile with redshift. For the calculations in Section 4 we fix $A=0.95$ and $\epsilon=0$, which, for our choice of halo properties and bright cluster removal (see below), corresponds to $\ell(\ell+1)C^{\rm tSZ}_{\ell}/2\pi|_{\ell=3000}=4~\mu$K$^2$ at 150~GHz. This is consistent with recent ACT and SPT measurements (\citeauthor{dunkley/etal:2011} 2011; R12). The effect of introducing departures from self-similar redshift-evolution in the electron pressure profile is discussed in Section 5.1.2.

\subsection{Halo definition}

We follow X12 in using the halo mass function from \cite{sheth/tormen:1999}, and the associated large-scale halo bias from \cite{sheth/etal:2001}. This mass function approximates the number density of haloes as a function of halo virial mass, $M_{\rm vir}$. The electron pressure profiles of \cite{arnaud/etal:2010} and the EM12 model are expressed not in terms of $M_{\rm vir}$, but $M_{500}$, defined as the mass of the region within which the mean overdensity is 500 times the critical background density at that redshift. When calculating $\tilde{y}_{\ell}$ we must therefore first calculate $M_{500}$ as a function of $M_{\rm vir}$ and redshift \citep[see, e.g., Section 2.1 of][]{komatsu/seljak:2002}.

Again following X12, we assume that the galaxy density profile, $u_{\rm gal}$, equals that of the dark matter in an NFW halo \citep{navarro/etal:1996}, $u_{\rm DM}$, and that haloes are truncated at the virial radius, $r_{\rm vir}$, for the purposes of hosting CIB sources. As in EM12, however, contributions to the Compton Y-parameter are allowed out to $4r_{\rm vir}$. The effect of halo truncation radius on the tSZ and clustered CIB power spectra is not negligible \citep[e.g.,][]{komatsu/seljak:2002,viero/etal:2009}, and it is not clear that these assumptions are reasonable over the range of source populations and redshifts considered. In Section 5.1.1 we consider the effect of forcing the CIB sources to lie at least some minimum distance from the centre of their host haloes, motivated by evidence that galaxies at the centers of groups and clusters are typically undergoing less-active star formation than those lying further out (e.g., \citeauthor{kennicutt:1983} 1983, \citeauthor{hashimoto/etal:1998} 1998, \citeauthor{bai/etal:2006} 2006, \citeauthor{bai/etal:2007} 2007; see also \citeauthor{boselli/gavazzi:2006} 2006 and references therein).

\subsection{Source removal}

Analyses of tSZ and CIB statistics are sensitive to the level of removal of bright resolved objects (clusters and point sources) from maps. Removal of objects down to the resolution limits of current instruments like \emph{Herschel}/SPIRE, \emph{Planck}, ACT and SPT has some effect on the tSZ one-halo power \citep[e.g.,][]{komatsu/kitayama:1999,shaw/etal:2009} and the CIB shot noise \citep[e.g.,][]{bethermin/etal:2011}, and will also affect the clustered CIB power to a lesser extent (which is, as yet, not quantified; see Section 3.1 of \citeauthor{addison/etal:2012} 2012).

As stated, X12 use the L11 CIB source model, accounting for removal of bright CIB sources by ignoring contributions to $\rmd I^{\rm CIB}/\rmd z$ from sources with flux above some threshold $S_{\rm cut}$, corresponding to the resolution limit of the relevant band (50~mJy for SPIRE bands, 6.4~mJy for the SPT 150~GHz band, and various values for \emph{Planck} -- see Table 3 of \citeauthor{planckcib:2011} 2011a). X12 also, somewhat arbitrarily, discount contributions to the clustered CIB power spectrum from $z<0.25$. We do likewise for the clustered CIB and tSZ$\times$CIB power; for the angular scales relevant for constraining the kSZ effect ($\ell\gtrsim3000$), the effect of this truncation is probably negligible, although this may not be the case on larger angular scales (see Section 4.3).

We remove the brightest 0.1 clusters per square degree in our model, roughly corresponding to a level of cluster detection currently achieved in deep ACT and SPT maps \citep{staniszewski/etal:2009,marriage/etal:2011,benson/etal:prep}. We do this to eliminate the effect of any massive, highly extended cluster at $z\simeq0$ (which, should it exist, would surely be masked in any real power spectrum analysis). The exact level of cluster removal has little effect on the tSZ$\times$CIB power, especially for $\ell\gtrsim3000$ (see Section 5.1.2). We define cluster brightness as the tSZ decrement integrated over the cluster's projected area. Due to effects such as the instrument beam and map filtering, this is not exactly the quantity that is measured in real maps, however, a more sophisticated treatment is beyond the scope of this work.

\section{Results}

In this section we use the equations of Section 2 and the tSZ and CIB models described in Section 3 to calculate several quantities: the redshift-dependence of the tSZ and CIB intensity, the angular power spectrum of the tSZ, clustered CIB and tSZ$\times$CIB components for a range of frequencies, the redshift-dependence of the tSZ, clustered CIB and tSZ$\times$CIB power, and the tSZ$\times$CIB correlation coefficient $\xi$ used by R12.

\subsection{Redshift-distribution of intensity, $\rmd I/\rmd z$}

Figure 1 shows the redshift-distribution of the tSZ and CIB intensity, plotted as $\rmd\ln I/\rmd z$. We obtained $\rmd\ln I^{\rm tSZ}/\rmd z$ using equation (17). In these units the tSZ intensity is independent of frequency. While the tSZ effect integrated over the projected area of a cluster is independent of the cluster redshift (neglecting redshift-evolution in halo concentration), the scarcity of massive haloes at high redshift results in $z<2$ clusters contributing the majority of the tSZ intensity.

The CIB intensity redshift-distribution is shown for two recent models (\citeauthor{bethermin/etal:2011} 2011 -- B11, and \citeauthor{lapi/etal:2011} 2011 -- L11) and two frequencies. Increasing $\nu$ and moving into the sub-mm results in an increase in the relative importance of low-redshift sources as well as an increase in the total CIB intensity (not shown in Figure 1). This behaviour is common to other recent CIB source models \citep[including][]{lagache/etal:2003,negrello/etal:2007,marsden/etal:2011}, and arises from the thermal dust emission from CIB sources resembling a modified blackbody peaked at a rest-frame wavelength of $\sim100~\mu$m. In the L11 model, the peak in $\rmd I^{\rm CIB}/\rmd z$ at $z\simeq1.5$ is due to starburst galaxies, whose contribution decreases rapidly at higher redshift such that for $z>2$ the proto-spheroids are the sole contributors. The structure in the $\rmd I^{\rm CIB}/\rmd z$ curves arises from summing the contributions from the spiral, starburst and proto-spheroid sources; the L11 1200 GHz curve in Figure 1 is the sum of the three 250 $\mu$m curves plotted in Figure 3 of X12.

\begin{figure}
	\centering
	\includegraphics{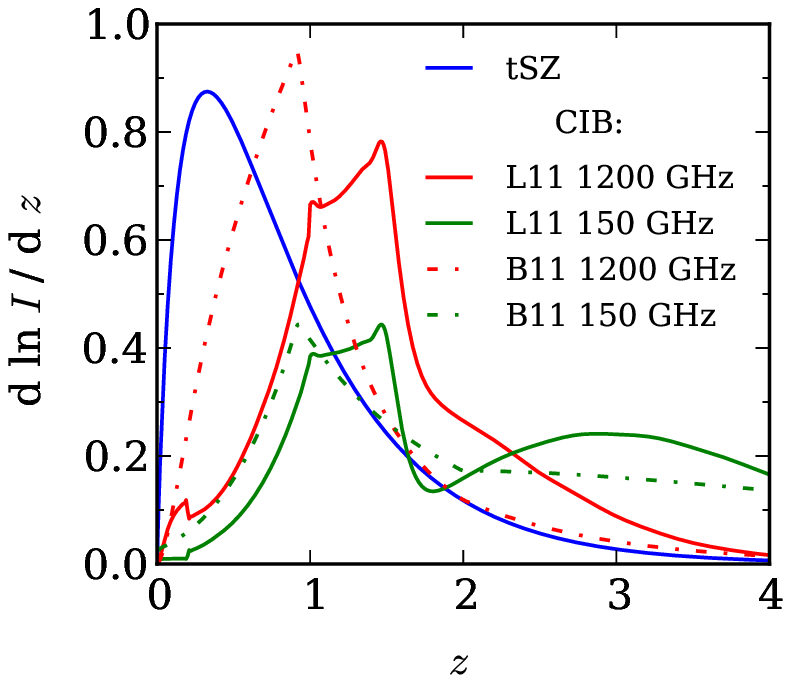}
	\caption{Redshift-distribution of tSZ and CIB intensity. The tSZ intensity distribution is calculated from the model of \protect\cite{efstathiou/migliaccio:2012} and, in these units, is independent of frequency. We show the CIB intensity distribution at 150 and 1200~GHz predicted by \protect\citeauthor{bethermin/etal:2011} (2011 -- B11) and \protect\citeauthor{lapi/etal:2011} (2011 -- L11). The overlap between the tSZ and CIB curves gives an indication of the size of the tSZ$\times$CIB cross-correlation. The degree of correlation increases for higher CIB frequencies because $\rmd I^{\rm CIB}/\rmd z$ becomes increasingly concentrated to low redshift.}
\end{figure}

The overlap between the tSZ and CIB curves on this plot give some indication of the size of the tSZ$\times$CIB cross-correlation, although, as shown in equations 26 and 27, the cross-power also depends on other factors, for instance the extent to which bright CIB sources occupy massive haloes for the one-halo term, and the effective redshift window function introduced by $P_{\rm DM}$ for the two-halo term.

Consider cross-correlating a map at a fixed frequency where the tSZ makes a significant contribution (e.g., 150~GHz) with maps at higher frequencies. Figure 1 suggests that the degree of tSZ$\times$CIB cross-correlation will \emph{increase} as the frequency of the second map increases (at least up to 1200~GHz). However, the degree of CIB$\times$CIB correlation will \emph{decrease} as $\rmd I^{\rm CIB}/\rmd z$ in the second map becomes more concentrated to low redshift. We therefore expect the importance of the tSZ$\times$CIB contribution relative to the cross-spectrum CIB power to increase as the separation of the two map frequencies increases. This is illustrated in Section 4.2, below.

The \cite{bethermin/etal:2011} CIB model predicts far more low-redshift flux than that of L11. The results in Section 4 follow X12 and use the L11 predictions, however we consider the effect of redistributing some of the high-redshift flux to lower redshift on the tSZ$\times$CIB power in Section 5.1.1.

\begin{figure*}
	\includegraphics{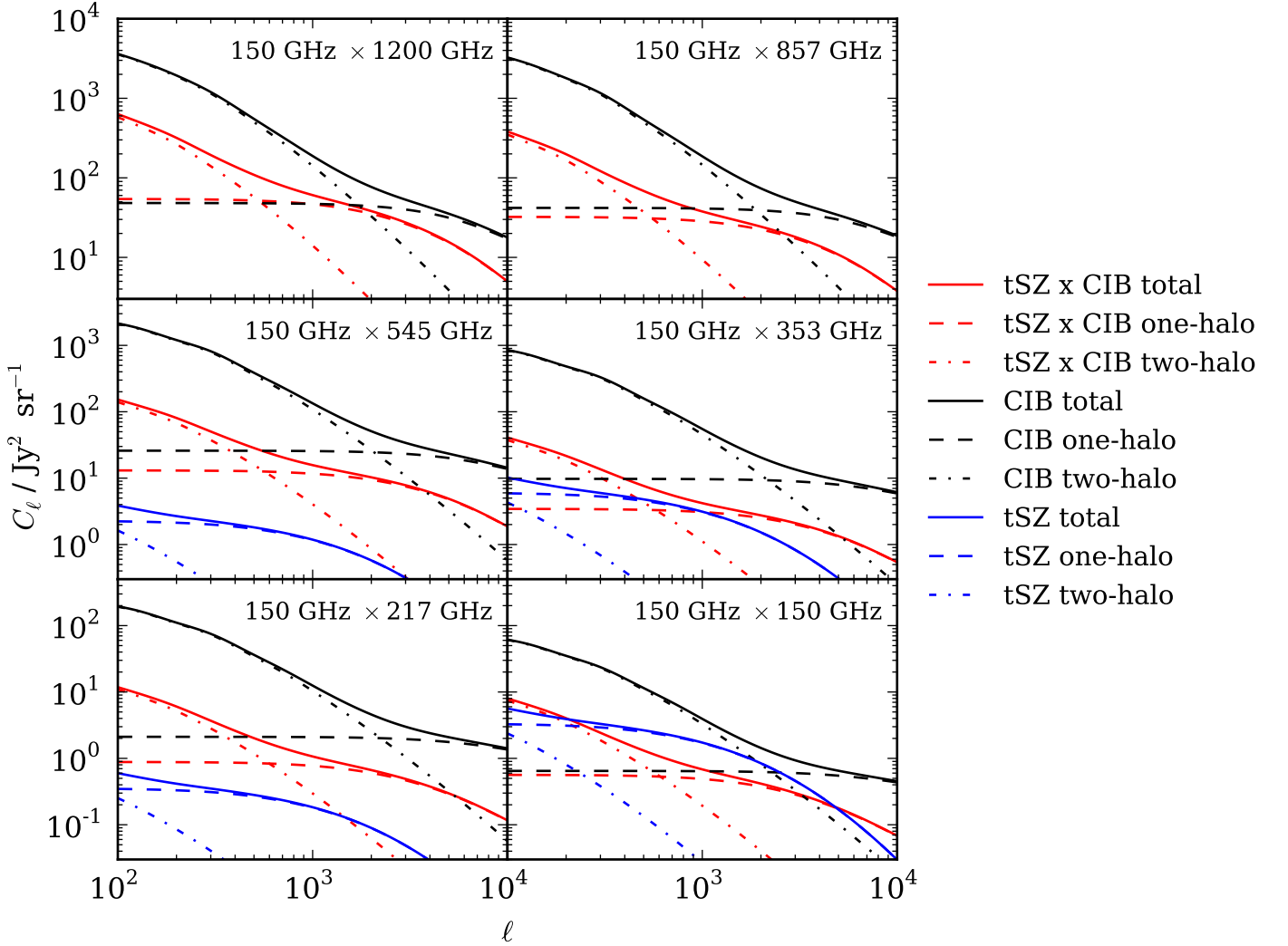}
	\caption{Angular power spectrum from the tSZ$\times$CIB correlation predicted for cross-correlating a map at 150~GHz with maps at various other frequencies using the model described in Sections 2 and 3. The CIB power here is clustered CIB power only (the Poisson shot noise power is not shown). Note that the tSZ$\times$CIB cross-power component is negative for each case shown here, while the tSZ power is negative in each case except $150\times150$~GHz. The importance of the tSZ$\times$CIB contribution increases relative to the CIB contribution as the frequency separation increases because the CIB becomes more concentrated to low redshift at higher frequencies, increasing the overlap with the tSZ clusters (see Figure 1). The tSZ$\times$CIB component is uncertain by a factor of two to three; to avoid cluttering the figure we do not show uncertainty estimates here, but see discussion in Section 5.1.}
\end{figure*}

\subsection{tSZ$\times$CIB angular power spectrum}

Figure 2 shows the clustered CIB, tSZ and tSZ$\times$CIB contributions to the angular power spectra from cross-correlating a 150~GHz map with maps at a range of higher frequencies. Note that the CIB Poisson shot noise is not shown because it is not yet well-constrained by data for the widely-spaced bands. Several features are apparent in these plots. Firstly, as expected, the size of the tSZ$\times$CIB power relative to the clustered CIB power generally increases as the frequency separation of the maps increases. The $150\times353$ and $150\times150~$GHz spectra are exceptions to this trend because both terms in parentheses in equations 26 and 27 are significant (in the other spectra, one of the two terms is highly sub-dominant). In the $150\times353~$GHz spectrum, the contribution from the tSZ in the 353~GHz map correlating with the CIB sources in the 150~GHz map acts to `fill in' around 10 per cent of the 150 tSZ $\times$ 353 CIB power, since the tSZ effect in the two maps has opposite signs. In the $150\times150~$GHz spectrum, the significant presence of both tSZ and CIB fluctuations in the map acts constructively and leads to a larger tSZ$\times$CIB component (relative to clustered CIB).

All three components shown in Figure 2 consist of one- and two-halo terms. As stated earlier, the scarcity of massive clusters means the two-halo tSZ power is sub-dominant over the whole range of angular scales shown (although masking more bright clusters would increase the relative importance of the two-halo term). For the X12 CIB clustering model, the tSZ$\times$CIB one-halo power dominates the two-halo power at significantly larger scales than for the clustered CIB. This is mainly because:
\begin{enumerate}
\item the low-redshift spirals and starbursts responsible for two thirds of the tSZ$\times$CIB cross-correlation power (compared to a third or less of the CIB power) are hosted in low-mass, and hence low-bias, haloes (reducing the two-halo power -- see equation 27), and 
\item there is no mechanism to limit the star formation or CIB source emissivity in massive haloes in the X12 model, possibly leading to an overestimation of the one-halo tSZ$\times$CIB power. 
\end{enumerate}
The relationship between CIB source emissivity and halo mass, and the modelling of the CIB clustering, are the largest sources of uncertainty in our tSZ$\times$CIB power calculation (see Section 5.1.1).

\begin{figure*}
	\includegraphics{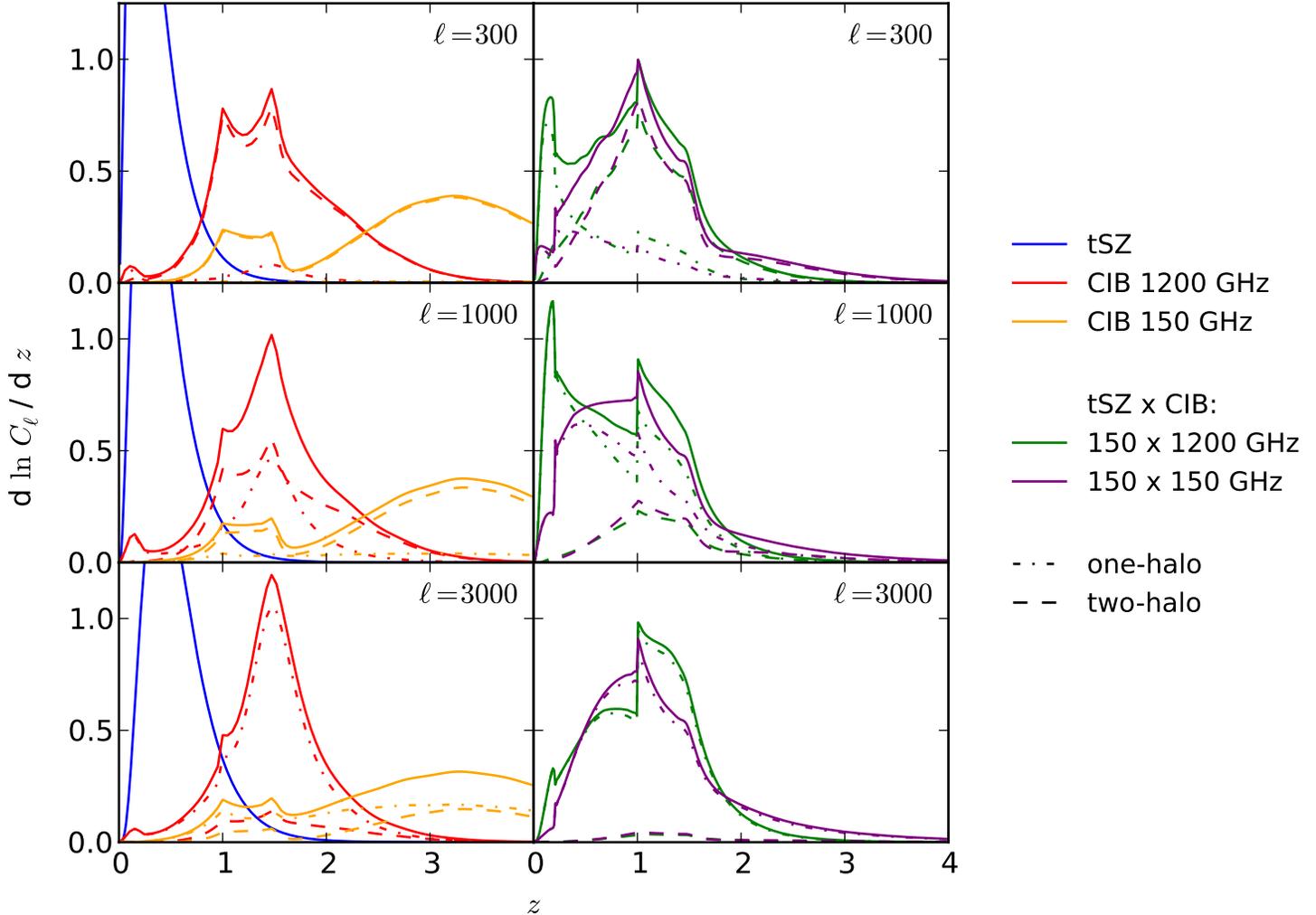}
	\caption{Redshift-distribution of the angular power spectrum, $\rmd \ln C_{\ell}/\rmd z$, for the tSZ and clustered CIB components (left panels; compare with $\rmd\ln I/\rmd z$ in Figure 1) and the tSZ$\times$CIB component (right panels). While the redshift-distribution of clustered CIB power changes significantly from 1200 to 150~GHz, that of the tSZ$\times$CIB power is fairly similar. The complex structure in $\rmd\ln C_{\ell}^{\rm tSZ\times CIB}/\rmd z$ is due to the existence of three different populations of CIB sources in our adopted CIB source model \protect\citep{lapi/etal:2011}. \protect\cite{xia/etal:2012} find that the clustering properties of the $z<1$ CIB sources, which make a significant contribution to the tSZ$\times$CIB power, are not constrained by angular power spectra of CIB fluctuations from \emph{Herschel}/SPIRE, \emph{Planck} and SPT, meaning that the low-$z$ tSZ$\times$CIB power is highly uncertain (see Section 5.1.1).}
\end{figure*}

\subsection{Redshift-distribution of power, $\rmd C_{\ell}/\rmd z$}

Figure 3 shows the redshift-distribution of power, plotted as $\rmd\ln C_{\ell}/\rmd z$, for the tSZ, clustered CIB and tSZ$\times$CIB components. Comparing the left-hand panels with Figure 1, the tSZ power and intensity have similar redshift distributions, as do the clustered CIB power and CIB intensity at 1200~GHz. The distribution of clustered CIB power at 150~GHz (particularly the two-halo power) is concentrated to higher redshifts than the intensity, with only a small contribution from the peak in $\rmd I/\rmd z$ at $z\la1$. In the X12 model, the proto-spheroid sources that contribute all the CIB at $z>2$ are hosted in massive haloes, which at these redshifts are highly biased (increasing the importance of these sources for contributing to the two-halo CIB power but not the intensity).

The redshift-distribution of the tSZ$\times$CIB power shows considerably less frequency dependence than that of the clustered CIB power. This is because most tSZ clusters lie at $z<2$ where, according to the \cite{lapi/etal:2011} model, the CIB sources are predominantly starbursts, and the shape of $\rmd I^{\rm CIB}/\rmd z$ is fairly similar for the range of frequencies considered here (Figure 1). The peak at $z\simeq0.15$ in the right-hand panels of Figure 3 is due to local spiral galaxies in the \cite{lapi/etal:2011} model. The sharp step at $z=1$ occurs because the high-redshift proto-spheroid population comes into existence discontinuously; while unrealistic, this feature does not affect our analysis because the power spectrum remains smooth and the redistribution of sources required to achieve a gradual transition in $\rmd C_{\ell}/\rmd z$ will induce a small change in the tSZ$\times$CIB power compared to other modelling uncertainties (see Section 5).

\begin{figure*}
	\includegraphics{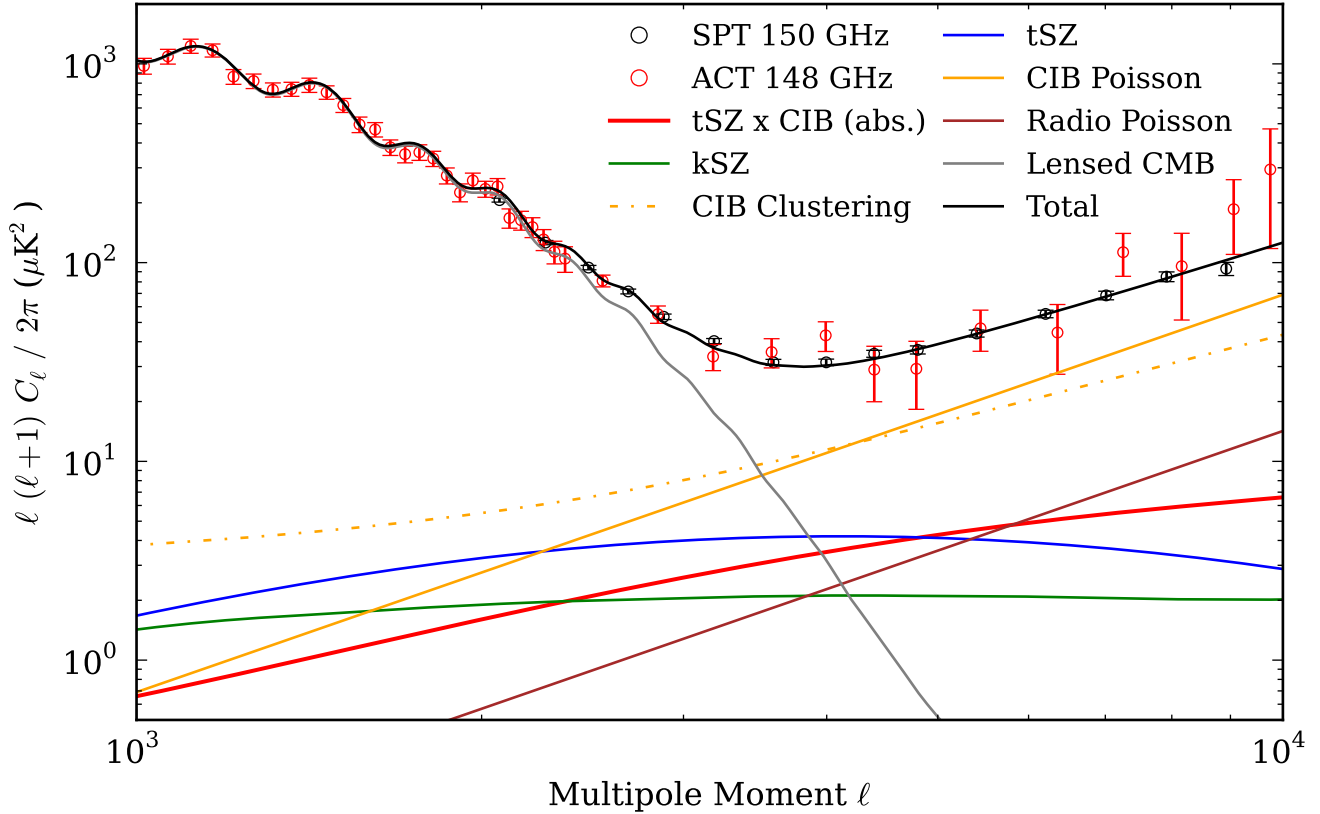}
	\caption{Comparison of power spectra of primary and secondary CMB temperature anisotropies and foregrounds at 150~GHz. The data points are the latest SPT (\protect\citeauthor{reichardt/etal:2012} 2011; R12) and ACT \protect\citep{das/etal:2011} measurements; we simply overplot the various power spectrum components here rather than performing a fit to these data. The CIB clustering power was reproduced from the model of \protect\citeauthor{xia/etal:2012} (2011; X12), as described in Section 3.1. The tSZ power spectrum was obtained from the model described in \protect\citeauthor{efstathiou/migliaccio:2012} (2012; EM12), fixed to have $\ell(\ell+1)C_{\ell}^{\rm tSZ}/2\pi|_{\ell=3000}=4~\mu$K$^2$ (see Section 3.2), and the tSZ$\times$CIB power, which is negative at 150~GHz, was calculated by combining the X12 and EM12 models, as described in Sections 2 and 3. We show the kSZ power calculated in \protect\cite{sehgal/etal:2010}. Radio and CIB point source shot noise levels were taken from R12 and X12 respectively (the ACT data points have been corrected to account for the difference in radio source shot noise levels due to more sources being masked by SPT). The primary lensed CMB power was calculated assuming a standard $\Lambda$CDM cosmology consistent with WMAP constraints \protect\citep{komatsu/etal:2011}.}
\end{figure*}

\subsection{Comparison to primary CMB and other foregrounds}

Figure 4 shows the most recent 150~GHz ACT and SPT power spectra (\citeauthor{das/etal:2011} 2011; R12), with contributions from primary and secondary CMB temperature anisotropies and foregrounds overplotted. The tSZ$\times$CIB power is a sub-dominant component but may be comparable to the kSZ at $\ell\gtrsim3000$. Note that the kSZ power shown in the figure includes the contribution from bulk electron motion in galaxy clusters and the intergalactic medium but assumes instantaneous reionization; including the effect of patchy reionization would increase this signal. Since the tSZ$\times$CIB power is negative for the principal CMB channels of ACT, SPT and \emph{Planck}, we would expect uncertainty in the tSZ$\times$CIB power to degrade constraints on the upper limit of the kSZ.

In principle, the tSZ$\times$CIB and kSZ components could be separated on the basis of their frequency dependence, however, we find that the frequency dependence is actually very similar across much of the frequency range probed by ACT and SPT. Figure 5 shows the frequency dependence of the tSZ, clustered CIB, tSZ$\times$CIB and kSZ power. The tSZ and clustered CIB power are -- individually -- easily distinguishable from a blackbody, however the tSZ$\times$CIB closely resembles a blackbody (horizontal line) for $\nu<200$~GHz. This will further worsen kSZ constraints, and indeed R12 find that the kSZ upper limit is increased by more than a factor of two when the tSZ$\times$CIB correlation is allowed, despite using data from all three SPT channels.

To assist in the analysis of small-scale CMB data, we have made the tSZ$\times$CIB curve from Figure 4 available to download\footnote{http://www.physics.ox.ac.uk/users/AddisonG/}.

\begin{figure*}
	\includegraphics{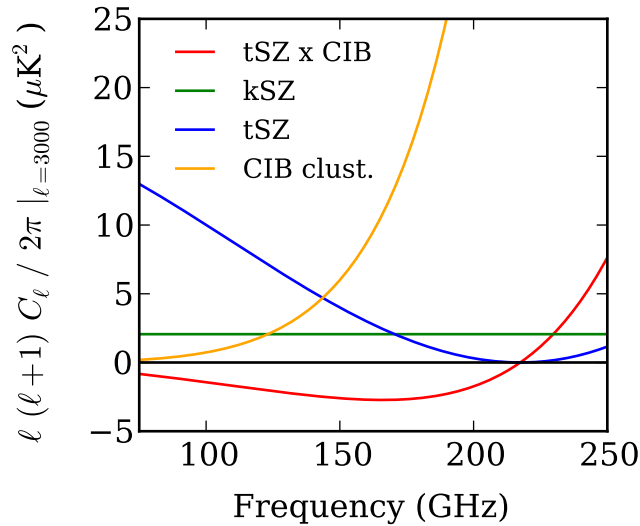}
	\caption{Frequency dependence of the tSZ$\times$CIB and kSZ power. The frequency dependence of the tSZ and clustered CIB power are shown for comparison; these signals are individually easy to distinguish from a blackbody but the tSZ$\times$CIB power has a frequency scaling that is very similar to that of the blackbody kSZ over the range of frequencies probed by ACT and SPT.}
\end{figure*}

\subsection{Correlation coefficient}

While, as mentioned, an advantage of the halo model approach is that there is no need for introducing a cross-correlation parameter into our model, quantifying the size of the tSZ$\times$CIB cross-correlation as a function of frequency and angular scale with such a parameter is worthwhile for the purposes of comparison with other models. R12 and \citealt{zahn/etal:2012} (2012, hereafter Z12) define a correlation coefficient, $\xi_{\ell}$, as
\be
\xi_{\ell}=\frac{C_{\ell,\nu_1\nu_2}^{\rm tSZ\times CIB}}{\sqrt{C_{\ell,\nu_1\nu_1}^{\rm tSZ}C_{\ell,\nu_2\nu_2}^{\rm CIB}}+\sqrt{C_{\ell,\nu_2\nu_2}^{\rm tSZ}C_{\ell,\nu_1\nu_1}^{\rm CIB}}}.
\ee
Note that $C_{\ell,\nu_1\nu_2}^{\rm CIB}$ here includes both the Poisson and clustered CIB components. When any $\ell$-dependence is neglected, R12 find $\xi=-0.18\pm0.12$ using SPT data at 95, 150 and 220 GHz. Z12 combine the tSZ model of \cite{shaw/etal:2010} with the various CIB source models presented in \cite{shang/etal:2012}, and find $-0.02>\xi_{3000}>-0.34$, with $\xi_{\ell}$ decreasing on larger angular scales (see Figure 3 in Z12).

Table 1 lists the values of $\xi_{\ell}$ at $\ell$ of 1000, 3000, 5000 and 9000 predicted from the X12 and EM12 models for cross-correlating a 150~GHz map with maps at the other frequencies shown in Figure 2. For the CIB Poisson shot noise levels, we follow X12 and use values of 6106, 4931, 1425, 175, 11 and 0.72 Jy$^2$ sr$^{-1}$ at 1200, 857, 545, 353, 217 and 150 GHz, respectively. We do not attempt to quantify the uncertainty in $\xi_{\ell}$ (but see discussion of power spectrum uncertainties in Section 5.1). The $\sim$20 per cent correlation obtained for 150$\times$150 and 150$\times$217~GHz is consistent with the findings of R12, and suggests that $\xi_{\ell}$ is not strongly dependent on frequency across the ACT and SPT bands. Physically this arises because the CIB is highly coherent at mm wavelengths, meaning that $C_{\ell}^{\rm tSZ\times CIB}$ and $\sqrt{C_{\ell}^{\rm CIB}}$ in equation (28) have a very similar frequency scaling.

It seems doubtful that a single correlation coefficient is sufficient to parametrize the tSZ$\times$CIB power for joint fitting to mm$\times$mm and mm$\times$sub-mm spectra. Table 1 shows an increase in $\xi_{\ell}$ of up to a factor of two for 150$\times$1200~GHz compared to 150$\times$150~GHz, which arises because the CIB redshift-distribution changes, becoming more concentrated to low-redshift, in the sub-mm bands (Figures 1 and 3).

Including the Poisson, as well as clustered, CIB power in equation (28) is unsatisfactory since rare, bright CIB sources may make a significant contribution to the Poisson power but be almost wholly irrelevant for the tSZ$\times$CIB. This could be the case in the mm bands if, for instance, much of the Poisson power comes from the luminous proto-spheroids predominantly in $M\sim10^{12.5-13}$~M$_{\sun}$ haloes at $z\gtrsim1.5$ \citep[as in][]{lapi/etal:2011}. Table 2 shows our predictions for $\xi_{\ell}$ if we take $C_{\ell}^{\rm CIB}$ in equation (28) to be the clustered CIB power only, as in \cite{shirokoff/etal:2011}. Unfortunately, due to the different $\ell$-dependence of the tSZ, clustered CIB and tSZ$\times$CIB power, this leads to an increase in the $\ell$-dependence of $\xi_{\ell}$.

Our results suggest that the frequency and angular scale dependence of $\xi$ may not be negligible, although further work is required to assess the extent to which fitting for a single correlation coefficient may bias other parameters.

\begin{table*}
  \centering
  \caption{Degree of tSZ$\times$CIB cross-correlation, $\xi_{\ell}$, for cross-correlating a 150~GHz map with maps at various frequencies}
  \begin{tabular}{lcccccc}
\hline
\multicolumn{1}{c}{$\ell$}&\multicolumn{1}{c}{SPIRE}&\multicolumn{1}{c}{\emph{Planck}}&\multicolumn{1}{c}{\emph{Planck}}&\multicolumn{1}{c}{\emph{Planck}}&\multicolumn{1}{c}{\emph{Planck}}&\multicolumn{1}{c}{150~GHz$^a$}\\
&1200~GHz&857~GHz&545~GHz&353~GHz&217~GHz&\\
\hline
1000&-0.27&-0.20&-0.14&-0.09&-0.11&-0.12\\
3000&-0.34&-0.26&-0.20&-0.14&-0.15&-0.17\\
5000&-0.35&-0.28&-0.22&-0.16&-0.16&-0.18\\
9000&-0.33&-0.28&-0.24&-0.17&-0.17&-0.18\\
\hline
\end{tabular}
\begin{center}
$^a$calculated with SPT 150~GHz bandpass filter; multiply values in this column by 1.1 or 1.15 for ACT 148 and \emph{Planck} 143~GHz bands respectively
\end{center}
\end{table*}

\begin{table*}
  \centering
  \caption{As Table 1 but with excluding the Poisson shot noise CIB power in equation (28) -- considering only clustered CIB}
  \begin{tabular}{lcccccc}
\hline
\multicolumn{1}{c}{$\ell$}&\multicolumn{1}{c}{SPIRE}&\multicolumn{1}{c}{\emph{Planck}}&\multicolumn{1}{c}{\emph{Planck}}&\multicolumn{1}{c}{\emph{Planck}}&\multicolumn{1}{c}{\emph{Planck}}&\multicolumn{1}{c}{150~GHz}\\
&1200~GHz&857~GHz&545~GHz&353~GHz&217~GHz\\
\hline
1000&-0.30&-0.22&-0.15&-0.10&-0.12&-0.13\\
3000&-0.44&-0.36&-0.27&-0.18&-0.22&-0.23\\
5000&-0.51&-0.43&-0.33&-0.23&-0.26&-0.26\\
9000&-0.60&-0.52&-0.40&-0.28&-0.30&-0.29\\
\hline
\end{tabular}
\end{table*}

\section{Discussion}

\subsection{Sources of uncertainty in tSZ$\times$CIB power}

Numerous sources of uncertainty in the tSZ$\times$CIB power are discussed below. Overall, we find that the contribution to the total tSZ$\times$CIB amplitude uncertainty is around a factor of two from uncertainties in the CIB modelling, tens of per cent from the tSZ treatment, and 10 per cent or less from other uncertainties.

\subsubsection{CIB modelling}

Two key assumptions in the X12 model are that (i) there is no connection between halo mass and CIB source spectral properties, and (ii) the clustering of CIB sources can be described by the following parameters: \{$M_{\rm min}$, $M_{\rm sat}$, $\sigma(\log_{10}M)$,$\alpha_{\rm sat}$\}, the parameter space that has been used to perform fits to angular correlation function measurements of (for instance) LRGs and SDSS galaxies \citep{zheng/etal:2005,zehavi/etal:2011}. What kind of uncertainty could be introduced in our tSZ$\times$CIB calculation if these assumptions are incorrect? This is a difficult question to answer without re-fitting to the angular power spectra used by X12 with different assumptions and parameters. Here we attempt to quantify the uncertainty with several simple examples, focussing on the 150$\times$150~GHz spectrum on angular scales of $\ell\simeq3000$, most relevant for attempts to constrain the kSZ power.

Firstly, we note that 60 per cent of the tSZ$\times$CIB power comes from haloes of mass over $10^{14}~M_{\sun}$, compared to only 10 per cent of the clustered CIB$\times$CIB. If star formation and far-infrared dust emission in these massive haloes is strongly suppressed compared to in less dense environments then the one-halo tSZ$\times$CIB power could be lower by a factor of two with little impact on the clustered CIB power. 

Secondly, X12 find $\alpha_{\rm sat}=1.81\pm0.04$ for the proto-spheroid sources that dominate the clustered CIB power spectra; a preference for $\alpha_{\rm sat}>1$ has also been reported in previous (similar) CIB clustering analyses \citep[e.g.,][]{amblard/etal:2011,planckcib:2011}, corresponding to the number of sources occupying a massive halo somehow increasing faster than the mass of the halo. This relation is not generally supported by simulations or analyses of clustering properties of other source populations, which favour $\alpha_{\rm sat}\la1$ \citep[e.g.,][]{kravtsov/etal:2004,zheng/etal:2005,reid/spergel:2009,white/etal:2011,zehavi/etal:2011}. We consider the possibility that the HOD parameters used for the CIB analyses, while capable of yielding a good fit to the angular power spectra, do not actually describe the true source properties, and that in fact $\alpha_{\rm sat}\simeq1$. Setting $\alpha_{\rm sat}=1$ causes the clustered CIB power to \emph{increase} by $\sim$30 per cent, but the tSZ$\times$CIB power to \emph{decrease} by 20 per cent, because reducing $\alpha_{\rm sat}$ for fixed $\rmd I^{\rm CIB}/\rmd z$ leads to more of the CIB contribution (both intensity and power) coming from low-mass haloes. This suggests that an inaccurate or incomplete HOD model may, potentially, be causing an overestimate of the tSZ$\times$CIB at the tens of percent level.

The results in Section 4 assume that CIB sources trace the dark matter density distribution in their host haloes within one virial radius (with the first, `central', source sitting at the very centre of the halo), as in X12. If we force all the CIB sources to lie at radii $r$ satisfying $0.5<r/r_{\rm vir}<1.0$ (though still tracing the dark matter), the clustered CIB$\times$CIB power at $\ell=3000$ decreases by only a few percent compared to 16 per cent for the tSZ$\times$CIB power. With a more extreme exclusion of CIB sources from the central parts of haloes, such that $1.0<r/r_{\rm vir}<2.0$, the clustered CIB$\times$CIB power decreases by 15 per cent and the tSZ$\times$CIB power by almost 60 per cent.

As shown in Figure 1, there is significant scatter in the low-redshift CIB flux predicted by current CIB source models. We find that increasing the L11 low-redshift spiral and starburst flux by 50 per cent (with a rescaling of the proto-spheroid flux such that the 150~GHz clustered CIB power at $\ell=3000$ is unchanged) leads to a 30 per cent increase in the tSZ$\times$CIB power. If the low-redshift source flux contribution is instead doubled (again with a rescaling of the proto-spheroid flux), the tSZ$\times$CIB power increases by some 60 per cent.

We finally note that X12 treat the spiral and starburst sources (SS) and the proto-spheroid (PS) sources completely independently. In fact, since both populations exist for $1<z<2$ (see Figure 3 of X12), with a significant overlap mass-wise in their host haloes, there should also be SS$\times$PS contributions to both the one- and two-halo clustered CIB power, which X12 do not include in their fitting.

\subsubsection{tSZ modelling and data}

The amplitude of the measured tSZ power spectrum is currently uncertain at the 10--30 per cent level, depending on what is assumed about, for instance, the kSZ component \citep[R12]{dunkley/etal:2011}. This corresponds to a smaller uncertainty in the amplitude of the tSZ$\times$CIB power, which is subdominant to the CIB modelling uncertainty discussed above.

As stated, the parameter $\epsilon$ in the EM12 model describes departure from self-similar redshift evolution of the cluster electron pressure profile. \cite{planck/earlyX:2011} find $\epsilon=0.66\pm0.52$ for a combined sample of clusters from \emph{Planck} and the Meta-Catalogue of X-ray detected Clusters (MCXC) at $z<1$ (see their equation 8 and Table 6, and also Section 2 of EM12). Holding the amplitude of the tSZ power at $\ell=3000$ fixed, we find a reduction in $C_{\ell=3000}^{\rm tSZ\times CIB}$ of $\sim$15 and 30 per cent for $\epsilon=0.5$ and 1.0 respectively, compared to the $\epsilon=0$ case. Note that holding $C^{\rm tSZ}_{\ell=3000}$ fixed requires setting the amplitude parameter $A=1.24$ and 1.58 for $\epsilon=0.5$ and 1.0 respectively. This redshift-evolution uncertainty is probably the largest source of uncertainty in our tSZ$\times$CIB model after the CIB modelling discussed above.

The results in Section 4 are shown with the brightest 0.1 deg$^{-2}$ clusters removed. For our adopted cosmology and halo definitions, this level of removal mainly affects power on large angular scales, with a negligible effect in the tSZ and tSZ$\times$CIB power at $\ell\ga3000$ \citep[consistent with the findings of][]{shaw/etal:2009}. We investigate the effect of removing ten times more clusters (1.0 deg$^{-2}$) and find that, while the tSZ$\times$tSZ power is reduced by almost a factor of two at $\ell=3000$, $C_{\ell}^{\rm tSZ\times CIB}$ decreases by only $\sim$10 per cent. As might be expected, removing more bright clusters increases the size of the tSZ$\times$CIB power compared to the tSZ.

\subsubsection{Other uncertainties}
\begin{enumerate}
\item Relativistic corrections and bulk motion of cluster electrons are expected to modify the tSZ power by $\sim$10 per cent \citep[e.g.,][]{nozawa/etal:1998,nozawa/etal:2006}, and will have a reduced effect on the tSZ$\times$CIB power.
\item Statistical uncertainty in proto-spheroid clustering parameters from X12 contributes a 10 per cent uncertainty at $\ell=3000$ in the clustered CIB power (inducing an uncertainty of $\sim$5 per cent in the tSZ$\times$CIB power), mainly due to the measured power spectra becoming dominated by shot noise on small scales. This measurement uncertainty is currently significantly smaller than the CIB modelling uncertainty discussed above.
\item Uncertainty in $\Lambda$CDM cosmological parameters: the tSZ effect depends very strongly on $\sigma_8$ -- $C_{\ell=3000}^{\rm tSZ}\propto\sigma_8^{\sim8}$ in our model ($C_{\ell=3000}^{\rm tSZ\times CIB}\propto\sigma_8^{\sim 6}$); we find that changing $\sigma_8$ by 0.03 \citep[corresponding to the 1$\sigma$ WMAP-7 uncertainty;][]{komatsu/etal:2011} leads to a $\sim$20 per cent change in the tSZ$\times$CIB amplitude, however, if the tSZ amplitude is held fixed, $C_{\ell=3000}^{\rm tSZ\times CIB}$ changes by $<10$ per cent. If a standard cosmology is assumed, uncertainty in cosmological parameters is not currently a dominant source of tSZ$\times$CIB uncertainty.
\item We have not investigated varying halo properties (e.g., mass function, concentration, or introducing scale-dependent or stochastic halo bias). We postpone assessing the impact of uncertainties in these quantities on the tSZ$\times$CIB power to future work given the size of the CIB modelling uncertainties discussed above.
\end{enumerate}

\subsection{Future data}

\subsubsection{Angular power spectra from mm and sub-mm maps}

Our results show that the contribution of the tSZ$\times$CIB power relative to the clustered CIB plus tSZ power increases for cross-correlating mm and sub-mm maps with widely-spaced frequencies (Figure 2). Cross-correlating ACT and Balloon-borne Large-Aperture Submillimeter Telescope (BLAST; observations at 600, 860 and 1200~GHz) maps of an 8.6 deg$^2$ area of sky near the south ecliptic pole revealed a significant detection of cross-spectra clustered CIB power \citep{hajian/etal:2012}, however the small sky area means these data are unable to provide constraints on the tSZ$\times$CIB power. Imminent \emph{Herschel}/SPIRE observations, also at 600, 860 and 1200 GHz, will provide $\sim$100 deg$^2$ of overlap with existing ACT\footnote{http://herschel.esac.esa.int/Docs/AO2/GT2\_accepted.html\#GT2\_mviero\_1}\footnote{http://herschel.esac.esa.int/Docs/AO2/OT2\_accepted.html\#OT2\_mviero\_2} and SPT\footnote{http://herschel.esac.esa.int/Docs/AO1/OT1\_accepted.html\#OT1\_jcarls01\_3} fields (as well as upcoming ACTPol -- \citeauthor{niemack/etal:2010} 2010 -- and SPTpol -- \citeauthor{mcmahon/etal:2009} 2009 -- fields), which should allow us to improve constraints on the CIB and tSZ$\times$CIB correlations.

In Section 4 we showed results calculated using the SPT 150 GHz bandpass filter. Since the CIB intensity and clustered power amplitude are falling off very steeply with decreasing frequency in the mm-bands \citep[e.g.][]{fixsen/etal:1998,planckcib:2011}, the contribution to the power spectrum from clustered CIB sources is $\sim$20 per cent lower in the ACT 148 GHz channel \citep{addison/etal:2012}. This means that the tSZ$\times$CIB power in the ACT 148~GHz $\times$ SPIRE cross-spectra will be around 10 per cent larger relative to the CIB$\times$CIB contribution compared to in SPT 150~GHz $\times$ SPIRE. The SPT 95~GHz channel is also potentially well-suited to constraining the tSZ$\times$CIB signal; the ratio of the tSZ$\times$CIB to the CIB power in SPT 95~GHz $\times$ SPIRE spectra will be some five times larger than for SPT 150~GHz $\times$ SPIRE, based on the frequency scaling of the tSZ and clustered CIB (\citeauthor{addison/etal:2012} 2012; R12). Cross-correlations of mm and sub-mm \emph{Planck} High- and Low-Frequency Instrument channels (e.g., 70$\times$857~GHz, 100$\times$857~GHz) may also be of use for constraining the tSZ$\times$CIB power for $\ell\lesssim2500$. Which channels give the best constraints in practice will also depend on the amount of noise in the maps and the size of the Poisson CIB power.

\subsubsection{Other statistics}

The angular power spectrum of mm and sub-mm maps is not the only statistic with the ability to explore the connection between clusters and CIB sources. Other measurements that may also provide insight include:
\begin{enumerate}
\item cross-correlations of CIB-dominated maps with X-ray cluster maps (either stacking analyses or power spectrum based),
\item targeted imaging of clusters in sub-mm bands (using, e.g., \emph{Herschel}, Submillimetre Common-User Bolometer Array-2 and the Atacama Large Millimeter/sub-millimeter Array) -- this will improve constraints on the spatial distribution of CIB sources within massive haloes, which enters into our modelling through $u_{\rm gal}$, and
\item cross-correlations of CIB maps with catalogues of objects at well-measured redshifts (e.g., LRGs, quasars) -- this will improve constraints on $\rmd I^{\rm CIB}/\rmd z$, particularly at $z<1$, most relevant for the tSZ$\times$CIB power; ACT, \emph{Herschel}, ACTPol and \emph{Planck} (as well as, e.g., IRAS at higher frequencies) have -- or will have -- the overlap with optical surveys required for these studies.
\end{enumerate}

\section{Conclusions}

We have presented equations describing the contribution to the angular power spectrum from the correlation of tSZ clusters and CIB sources in a halo model framework. We then used these equations to calculate the tSZ$\times$CIB power spectrum at 150~GHz and for a range of cross-spectra using recent tSZ and CIB halo models. We find that:
\begin{enumerate}
\item{The tSZ$\times$CIB is a sub-dominant component of the angular power spectrum at 150~GHz, contributing approximately $-2$~$\mu$K$^2$ at $\ell=3000$; uncertainty in the tSZ$\times$CIB power will degrade kSZ constraints, as found in recent SPT analysis (R12, Z12), due to the similarity in their frequency dependence.}
\item{The size of the tSZ$\times$CIB power relative to the clustered CIB power increases if we correlate mm and sub-mm maps with increasing frequency separation (at least up to 1200~GHz, the highest frequency \emph{Herschel}/SPIRE channel).}
\item{Uncertainty in the amplitude of the tSZ$\times$CIB power is currently a factor of two or three, with the uncertainty dominated by uncertainty in modelling of the clustering of CIB sources (although there are other significant sources of uncertainty, including possible evolution in cluster electron pressure profiles with redshift).}
\end{enumerate}

The framework laid out in Section 2 may be used to perform joint fitting to ACT, SPT, \emph{Herschel}, \emph{Planck} and other data sets. The halo model and HOD formalisms are not without their complications, and work remains in order to find the best way to parameterize the connection between CIB sources and their haloes; it seems essential that a thorough exploration of modelling assumptions accompany future analysis.

Detection of the tSZ$\times$CIB power in mm$\times$sub-mm cross-spectra may enable us to constrain the tSZ effect from unresolved clusters in spectra that are free from the signal of the primary CMB (although the CMB fluctuations will still exist as noise in the mm maps). Whether this can lead to improved tSZ constraints over mm-band data alone will depend on the true size of the tSZ$\times$CIB signal and future improvements in the modelling of the CIB sources.

We finally remark that the tSZ$\times$CIB power may be a significant contaminant for detection of signals beyond the kSZ, for example, detection of the Integrated Sachs-Wolfe effect by cross-correlating mm-band CMB maps with tSZ or CIB maps \citep{taburet/etal:2011,ilic/etal:2011}.

\setlength{\parskip}{2ex}

The authors are indebted to Jun-Qing Xia, Andrea Lapi and Marina Migliaccio for correspondence regarding the models described in Section 3. We also thank Christian Reichardt and Oliver Zahn for helpful discussions, Eiichiro Komatsu for reading the manuscript and making useful suggestions, Ryan Keisler for comments regarding the lensing of the CIB sources, and the referee for his or her comments. GA is supported by an STFC studentship, and JD is recipient of an RCUK Fellowship and ERC grant 259505.

\end{document}